\documentclass[11pt]{article}
\usepackage{latexsym,graphicx,amssymb, amsmath}

 \def\pd{\partial} \def\pp{\prime}  \def\b{\beta} \def\dl{\delta} \def\s{\sigma}  \def\eps{\epsilon}  \def\lam{\lambda} \def\Lam{\Lambda}  \def\Gm{\Gamma} \def\om{\omega}  \def\sq{\sqrt} \def\fr{\frac} \def\half{\frac{1}{2}}

 \def\hg{{\hat g}} \def\bg{{\bar g}}  \def\nb{\nabla}

\def\lap3{~| \!\!\! \partial^2} \def\dlap3{~| \!\!\! \partial^4} \def\invlap3{~| \!\!\! \partial^{-2}}
\def\lang{\langle} \def\rang{\rangle}
\def\d3x{d^3 {\bf x}~}

\def\geh{\mathfrak{g}}
\def\QCD{{\rm QCD}}

\begin{document}

\begin{flushright}
October 2020
\end{flushright}

\begin{center}
{\large {\bf Diffeomorphism Invariance Demands Conformal Anomalies}} 
\end{center}

\begin{center}
{\sc Ken-ji Hamada}
\end{center}

\begin{center}
{\it Institute of Particle and Nuclear Studies, KEK, Tsukuba 305-0801, Japan  \\ and
Department of Particle and Nuclear Physics, The Graduate University for Advanced Studies (SOKENDAI), Tsukuba 305-0801, Japan}
\end{center}

\begin{abstract}
We study a series of the Wess-Zumino actions obtained by repeatedly integrating conformal anomalies with respect to the conformal-factor field that appear at higher loops. We show that they arise as physical quantities required to make nonlocal loop correction terms diffeomorphism invariant. Specifically, in a conformally flat spacetime $ds^2=e^{2\phi}(-d\eta^2 + d{\bf x}^2)$, we find that effective actions are described in terms of momentum squared expressed as a physical $Q^2 = q^2/e^{2\phi}$ for $q^2$ measured by the flat metric, which recalls the relationship between physical momentum and comoving momentum in cosmology. It is confirmed by calculating the effective action of QED in such a curved spacetime at the 3-loop level using dimensional regularization. The same applies to the case of QCD, in which we show that the effective action can be summarized in the form of the reciprocal of a running coupling constant squared described by the physical momentum. We also see that the same holds for renormalizable quantum conformal gravity and that conformal anomalies are indispensable for formulating the theory.
\end{abstract}

\vspace{3mm}

\section{Introduction}
\setcounter{equation}{0}
\noindent

In general, when a symmetry that holds classically is broken by quantum effects, it is called an anomaly. Conformal anomalies imply that even if a classical action has conformal symmetry, it breaks down at the quantum level \cite{cd, ddi, duff, acd, bc, hathrellS, hathrellQED, freeman, ds, hamada14CS}. Hence, letting $\Gm$ be an effective action of the theory, a conformal variation of $\Gm$ is a conformal anomaly. That is, the trace of the energy-momentum tensor which vanishes classically becomes nonzero by quantum effects. For this reason, conformal anomalies are also called trace anomalies.

This paper will focus on the role of  conformal anomalies. As the name implies, conformal anomalies break conformal invariance, but it does not mean that they are anomalous. Rather, we will see that they are physical  quantities required to preserve diffeomorphism invariance, and are associated with nonlocality caused by quantization.

The relation between conformal anomaly and diffeomorphism invariance can be seen as follows. Let $f$ be a field that has conformal invariance classically and $I$ be its action in curved spacetime. The line element is defined by $ds^2 = g_{\mu\nu} dx^\mu dx^\nu$ and the metric field is decomposed into a conformal factor and other modes  as 
\begin{eqnarray}
      g_{\mu\nu} =e^{2\phi}\bg_{\mu\nu} ,
          \label{conformal metric decomposition}
\end{eqnarray}
then the relation $I(f,g)=I(f,\bg)$ holds.\footnote{
Depending on the type of the field, it is necessary to rescale the field appropriately to exclude the $\phi$-dependence.} 
On the other hand, a path integral measure $[df]_g$ that is diffeomorphism invariant generally depends on the conformal-factor field $\phi$. Therefore, by extracting the $\phi$-dependence of the measure as $[df]_g = e^{iS(\phi,\bg)} [df]_\bg$, we rewrite the effective action in the form 
\begin{eqnarray*}
    e^{i \Gm(g)}   \!\!\!\!&=&\!\!\!\!    \int [df]_g  \, e^{iI(f,g)}
                     \nonumber \\
         \!\!\!\!&=&\!\!\!\!
              e^{i S(\phi, \bg)}  \! \int [df]_\bg  \, e^{iI(f,\bg)}  = e^{i S(\phi, \bg)} e^{i \Gm(\bg)} ,
\end{eqnarray*}
that is, $\Gm(g)=S(\phi,\bg)+\Gm(\bg)$.\footnote{
When calculating the effective action, normally a background field of $f$ is introduced.} 
Hence, a variation of $S$ by $\phi$ gives a conformal anomaly.

Now, let us consider a simultaneous transformation
\begin{eqnarray*}
   \phi \to \phi - \om,   \qquad \bg_{\mu\nu} \to e^{2\om} \bg_{\mu\nu}
\end{eqnarray*}
that does not change the full metric $g_{\mu\nu}$, that is, preserves diffeomorphism invariance. Applying it to the above expression of the effective action, the left-hand side is trivially invariant, while the right-hand side is
\begin{eqnarray*}
     e^{iS(\phi-\om,e^{2\om}\bg)} e^{i\Gm(e^{2\om}\bg)} 
     = e^{iS(\phi-\om,e^{2\om}\bg)} e^{iS(\om,\bg)} e^{i\Gm(\bg)} .
\end{eqnarray*}
In order for this expression to return to the original $e^{i\Gm(g)}$, $S$ must satisfy 
\begin{eqnarray}
     S(\phi-\om,e^{2\om}\bg) + S(\om,\bg) = S(\phi, \bg) .
            \label{wess-zumino condition}
\end{eqnarray}
This relation is called the Wess-Zumino consistency condition \cite{wz, bcr}, and thus $S$ is called the Wess-Zumino action. This is another expression of the integrability condition often represented by $[\dl_{\phi}, \dl_{\phi^\pp}] \Gm =0$, where $\dl_\phi$ is a conformal variation.

In the following sections, we will investigate the Wess-Zumino actions that arise at higher loops, mainly focusing on conformal anomalies related to gauge fields in curved spacetime, and see that they are quantities necessary to construct a diffeomorphism invariant effective action. In particular, we will  concretely calculate the effective action of QED at the 3-loop level employing dimensional regularization, and show that the $\phi$-dependence is actually involved in a physical momentum measured by the full metric $g_{\mu\nu}$. The same is true for QCD, or Yang-Mills theory. That also applies to quantum conformal gravity, and other series of the Wess-Zumino actions that consists only of the gravitational field will be discussed.

\section{Wess-Zumino Actions at Higher Loops}
\setcounter{equation}{0}
\noindent

It is known that there are various types of conformal anomalies \cite{cd, ddi, duff, acd, bc, hathrellS, hathrellQED, freeman, ds, hamada14CS}. We here consider three types: the field strength squared of gauge fields, the Weyl tensor squared, and the (modified) Euler density. These are collectively written as ${\cal L}^{(0)}(g)$, and its spacetime volume integral is denoted by $\Gm^{(0)}(g)=\int d^4 x \, {\cal L}^{(0)}(g)$, that gives a tree part of the effective action. The action  is integrable with respect to the conformal-factor field $\phi$, that is, $[ \dl_{\phi}, \dl_{ \phi^\pp}] \Gm^{(0)}(g) = 0$.

The first Wess-Zumino action that appears at the 1-loop level is obtained by integrating ${\cal L}^{(0)}(g)$ with respect to the conformal-factor field from $0$ to $\phi$, written as 
\begin{eqnarray}
      S^{(1)}(\phi, \bg) = \int d^4 x \int^\phi_0  \!\! d\phi  \, {\cal L}^{(0)}(g) .
             \label{first wess-zumino action}
\end{eqnarray}
It is obvious that this satisfies the Wess-Zumino consistency condition (\ref{wess-zumino condition}), which can be shown by decomposing the range of  the integration $[0, \phi]$ into $[0,\om]$ and $[\om, \phi]$, where
\begin{eqnarray*}
   \int d^4 x \int^\phi_\om  \!\!  d\phi  \, {\cal L}^{(0)}(g) 
    = \int d^4 x \int^{\phi-\om}_0  \!\!\!  d\phi^\pp  \, {\cal L}^{(0)}(e^{2\phi^\pp + 2\om} \bg)
   = S^{(1)}(\phi-\om, e^{2\om} \bg) .
\end{eqnarray*}
Note here that it is essential that the integrand ${\cal L}^{(0)}(g)$ is covariant, that is, composed of the full metric $g_{\mu\nu}$.

Next, we consider what kind of the Wess-Zumino action appears at higher loops. First, let us simply repeat the definite integral for $\phi$ performed in (\ref{first wess-zumino action}) $n$ times. Writing the resulting action as $S_{\rm top}^{(n)}(\phi,\bg)$, it satisfies 
\begin{eqnarray}
       \int d^4 x \fr{\dl}{\dl \phi(x)} S_{\rm top}^{(n)}(\phi, \bg) = S_{\rm top}^{(n-1)}(\phi, \bg) .
               \label{reduction formula for top term}
\end{eqnarray}
Here, $S_{\rm top}^{(1)}$ is the Wess-Zumino action $S^{(1)}$ that satisfies the consistency condition (\ref{wess-zumino condition}). However, others do not satisfy it, that is, they are not the Wess-Zumino actions, because the integrand $S^{(n-1)}_{\rm top}$ to derive $S^{(n)}_{\rm top}$ is no longer  a covariant function of the full metric $g_{\mu\nu}$ for $n >1$.

In order to find the Wess-Zumino action that arises at higher loops, we need to find a diffeomorphism invariant effective action $\Gm^{(1)}(g)=\int d^4 x \, {\cal L}^{(1)}(g)$, which can be expressed as 
\begin{eqnarray*}
    \Gm^{(1)}(g) = S^{(1)}(\phi,\bg) + \Gm^{(1)}(\bg) ,
\end{eqnarray*}
where $\Gm^{(1)}( \bg)$ is a loop correction of the theory defined on the metric $\bg_{\mu\nu}$, generally has a nonlocal form. By using  the covariant ${\cal L}^{(1)}(g)$ as the integrand of the $\phi$-integral, we can obtain the second Wess-Zumino action $S^{(2)}(\phi,\bg)$. Furthermore, adding a nonlocal $\Gm^{(2)}(\bg)$ to $S^{(2)}(\phi,\bg)$ appropriately to make a diffeomorphism invariant $\Gm^{(2)}(g)= \int d^4 x {\cal L}^{(2)}(g)$ and integrating ${\cal L}^{(2)}(g)$, we obtain $S^{(3)}(\phi, \bg)$. Repeating this procedure, we find
\begin{eqnarray*}
      S^{(n+1)} (\phi, \bg) = \int d^4 x \int^\phi_0  \!\! d\phi  \, {\cal L}^{(n)}(g) 
\end{eqnarray*}
for $\Gm^{(n)}= \int d^4 x \, {\cal L}^{(n)}(g)$. Here, $S^{(n)}$ contains $S_{\rm top}^{(n)}$ as the term with the highest power of $\phi$.

The diffeomorphism invariant effective action will then be given by using $\Gm^{(n)}(g)$ as
\begin{eqnarray*}
    \Gm (g) = \sum_{n=0}^\infty a_n \Gm^{(n)}(g) ,
\end{eqnarray*}
where the coefficient $a_n$ is a function of the coupling constant.

In the following, taking the case of gauge fields as an example, a specific expression of the Wess-Zumino action and the effective action will be given. If the Minkowski metric $\eta_{\mu\nu}=(-1,1,1,1)$ is employed as $\bg_{\mu\nu}$, then $\Gm(\bg=\eta)$ is the ordinary effective action. For simplicity, let $\phi$ be a constant and consider in momentum space. Still,  in this example, the covariant structure we want to see is well retained. As for conformal anomalies composed of the gravitational field, we will discuss in the penultimate section.

The tree action of gauge fields is conformally invariant in 4 dimensions and $\sq{-g}F^a_{\mu\nu} F^a_{\lam\s} g^{\mu\lam} g^{ \nu\s}=F^a_{\mu\nu} F^a_{\lam\s} \eta^{\mu\lam} \eta^{\nu\s}$ holds, where $F^a_{\mu\nu}$ is a field strength and $a$ is the index of gauge group. From this, ${\cal L}^{(0)}(q)$ is given by $F^a_{\mu\nu}F^a_{\lam\s} (q) \, \eta^{\mu\lam} \eta^{\nu\s}$, where $q$ represents momentum measured in Minkowski spacetime. Therefore,  we can easily perform the integral of $\phi$, and obtain $S^{(1)}(\phi, \eta)$ as 
\begin{eqnarray}
     S^{(1)}(\phi, \eta) = \int \! \fr{d^4 q}{(2\pi)^4} \,  \phi  \, 
              F^a_{\mu\nu}F^a_{\lam\s} (q)  \, \eta^{\mu\lam} \eta^{\nu\s} .
                \label{first wess-zumino action for gauge field}
\end{eqnarray}
Further performing the $\phi$-integration $n$ times gives 
\begin{eqnarray*}
    S_{\rm top}^{(n)}(\phi,\eta) = \fr{1}{n!} \int  \fr{d^4 q}{(2\pi)^4} \, \phi^n  
         F^a_{\mu\nu}F^a_{\lam\s} (q)  \, \eta^{\mu\lam} \eta^{\nu\s}  .
\end{eqnarray*}

A diffeomorphism invariant form of the effective action is obtained by adding a loop correction $\Gm^{(1)}(\eta)$, known to be given by a logarithmic function of momentum, to the first Wess-Zumino action (\ref{first wess-zumino action for gauge field}). It will be given by 
\begin{eqnarray*}
    \Gm^{(1)} (g)  \!\!\!\!&=&\!\!\!\! 
               \int \! \fr{d^4 q}{(2\pi)^4} \, \biggl[ \phi  
                - \half \log \biggl( \fr{q^2}{\mu^2} \biggr) \biggr] 
                F^a_{\mu\nu}F^a_{\lam\s} (q) \, \eta^{\mu\lam} \eta^{\nu\s} 
                      \nonumber \\
         \!\!\!\!&=&\!\!\!\!
                  - \half  \int \! \fr{d^4 q}{(2\pi)^4} \, \log \biggl( \fr{Q^2}{\mu^2} \biggr) 
                       F^a_{\mu\nu}F^a_{\lam\s} (q) \, \eta^{\mu\lam} \eta^{\nu\s} ,
\end{eqnarray*}
where $q^2=\eta^{\mu\nu}q_\mu q_\nu$ and it has been known that the logarithmic correction term arises exactly with this coefficient according to that of $\phi$. $Q$ is momentum measured by the full metric $g_{\mu\nu}$, that is,
\begin{eqnarray}
       Q^2 = g^{\mu\nu} q_\mu q_\nu = \fr{q^2}{e^{2\phi}} .
          \label{physical momentum}
\end{eqnarray}
In cosmology, $Q$ is often called physical momentum, while $q$ is called comoving momentum.

In gauge theories, the effective actions up to 2 loops can be written using $\Gm^{(1)}(g)$. At 3 loops or more, the second Wess-Zumino action will appear, which is given by
\begin{eqnarray*}
    S^{(2)}(\phi, \eta)   \!\!\!\!&=&\!\!\!\!
               \int \! \fr{d^4 q}{(2\pi)^4} \int^\phi_0  \!\!  d\phi  \, {\cal L}^{(1)}(g) 
                  \nonumber \\
       \!\!\!\!&=&\!\!\!\!
              \int \! \fr{d^4 q}{(2\pi)^4}  \, \biggl[ \half \phi^2  
             - \half \phi \log \biggl( \fr{q^2}{\mu^2} \biggr) \biggr] 
             F^a_{\mu\nu}F^a_{\lam\s} (q) \, \eta^{\mu\lam} \eta^{\nu\s} .
\end{eqnarray*}
Here, we can see that this action satisfies the Wess-Zumino consistency condition (\ref{wess-zumino condition}) using the fact that when changing the Minkowski metric as $\eta_{\mu\nu} \to e^{2\om} \eta_{\mu\nu}$, the momentum squared is changed to $q^2 \to e^{-2\om} q^2$.

Adding a nonlocal loop correction $\Gm^{(2)}(\eta)$ to the above gives 
\begin{eqnarray*}
    \Gm^{(2)}(g)  \!\!\!\!&=&\!\!\!\! 
              \int \! \fr{d^4 q}{(2\pi)^4}  \, \biggl[ \half \phi^2  
             - \half \phi \log \biggl( \fr{q^2}{\mu^2} \biggr) 
             + \fr{1}{8} \log^2 \biggl( \fr{q^2}{\mu^2} \biggr)   \biggr] 
              F^a_{\mu\nu}F^a_{\lam\s} (q) \, \eta^{\mu\lam} \eta^{\nu\s} 
                   \nonumber \\
      \!\!\!\!&=&\!\!\!\!
              \fr{1}{8} \int \! \fr{d^4 q}{(2\pi)^4} \, \log^2 \biggl( \fr{Q^2}{\mu^2} \biggr) 
               F^a_{\mu\nu}F^a_{\lam\s} (q) \, \eta^{\mu\lam} \eta^{\nu\s}   ,     
\end{eqnarray*}
where the last logarithm squared term in the first line is $\Gm^{(2)}(\eta)$, and its coefficient is chosen so that the momentum is in the form of $Q$. Further repeating this procedure gives
\begin{eqnarray}
    \Gm^{(n)}(g) = \fr{(-1)^n}{2^n n!}  \! \int \! \fr{d^4 q}{(2\pi)^4} \, 
              \log^n \biggl( \fr{Q^2}{\mu^2} \biggr) 
              F^a_{\mu\nu}F^a_{\lam\s} (q) \, \eta^{\mu\lam} \eta^{\nu\s} .
           \label{final diffeomorphism invariant form}
\end{eqnarray}
In the following, we will see that the effective action is yielded actually in this form.

\section{QED Effective Action and Diffeomorphism Invariance}
\setcounter{equation}{0}
\noindent

Here, we will calculate the effective action of massless QED in curved spacetime to confirm the form (\ref{final diffeomorphism invariant form}) at the 3-loop level. As a method to perform the renormalization calculations, dimensional regularization is employed \cite{cd, ddi, duff, acd, bc, hathrellS, hathrellQED, freeman, ds, hamada14CS, hamada02, hamada14, hm16, hm17, book}. This is the only known regularization method that can do high-loop calculations while preserving diffeomorphism invariance as well as gauge invariance.

The advantage of this method is that the result does not depend on how to choose the path integral measure. This property comes from the fact that in a 4-dimensional method such as the DeWitt-Schwinger method, conformal anomalies as contributions from the measure are derived by regularizing a divergent quantity $\dl^{(4)}(0)=\lang x | x^\pp \rang|_{x^\pp \to x}$, whereas in dimensional regularization such a quantity identically vanishes due to $\dl^D(0)=\int d^D q =0$, where $D$ is spacetime dimension. In dimensional regularization, conformal anomalies are hidden between $D$ and $4$ dimensions, which arise as finite quantities yielded by canceling poles of ultraviolet (UV) divergences with zeros representing deviations of $D$-dimensional actions  from 4 dimensions as
\begin{eqnarray*}
    \fr{1}{D-4} \times o(D-4) \to \hbox{finite}.
\end{eqnarray*}

First, we summarize some facts concerning with renormalization group (RG) equations. Let $e_0$ be a bare coupling constant of QED, $e$ be its dimensionless renormalized coupling constant, and $Z_e$ be a renormalization factor connecting them. Let $Z_3^{1/2}$ be a renormalization factor of the gauge field, then the Ward-Takahashi identity $Z_e Z_3^{1/2}=1$ holds, so that $e_0 = \mu^{2-D/2} Z_3^{-1/2} e$, where $\mu$ is an arbitrary mass scale introduced to make up for the missing dimension. Since bare quantities do not depend on the scale $\mu$, we obtain
\begin{eqnarray*}
   \mu \fr{d e_0}{d\mu} = \mu^{2-D/2} Z_3^{-1/2} \biggl[ - \half (D-4) e 
       - \half e \mu \fr{d}{d\mu} \log Z_3 + \mu \fr{de}{d\mu} \biggl] = 0 .
\end{eqnarray*}
Thus, writing as $\bar{\b}_e(e) = (e/2) \mu d (\log Z_3)/d\mu$, the beta function can be expressed as
\begin{eqnarray*}
    \b_e (e, D) = \mu \fr{d e}{d\mu} = \half (D-4) e + \bar{\b}_e (e) .    
\end{eqnarray*}
If the renormalization factor is Laurent-expanded like $\log Z_3 = \sum_{n=1}^\infty f_n(e)/(D-4)^n$, then the beta function is expanded as 
\begin{eqnarray*}
   {\bar \b}_e(e) 
   = \fr{e}{2} \biggl\{  \fr{e}{2} \fr{\pd f_1(e)}{\pd e} 
            + \sum_{n=1}^\infty  \fr{1}{(D-4)^n} \biggl[ \fr{e}{2} \fr{\pd f_{n+1}(e)}{\pd e} 
               + \bar{\b}_e(e)  \fr{\pd f_n(e)}{\pd e} \biggr] \biggr\} ,
\end{eqnarray*}
where $\mu d/d\mu = (\mu de/d\mu ) \, \pd/\pd e = \b_e(e,D) \, \pd /\pd e$ is used. Since this quantity is finite and the pole terms must disappear, we find that the beta function can be expressed using  the simple-pole residue as
\begin{eqnarray}
    \bar{\b}_e(e) = \fr{e^2}{4} \fr{\pd f_1(e)}{\pd e} 
         \label{beta function using f1}
\end{eqnarray}
and the residues satisfy a RG equation
\begin{eqnarray}
      \fr{\pd f_{n+1}(e)}{\pd e} + \fr{2 \bar{\b}_e(e)}{e}  \fr{\pd f_n(e)}{\pd e} = 0 
          \label{renormalization group for fn}
\end{eqnarray}
for $n \geq 1$.

Here, the expansion of the beta function is expressed as 
\begin{eqnarray*}
    \bar{\b}_e(e) = b_1 e^3 + b_2 e^5 + b_3 e^7 +\cdots .
\end{eqnarray*}
Specific values of the coefficients are given by $b_1 = 4/3(4\pi)^2$, $b_2 = 4/(4\pi)^4$, and $b_3=-62/9(4\pi)^6$ \cite{gkls}, but the following discussion proceeds without using these values. For this expression, solving (\ref{beta function using f1}) and (\ref{renormalization group for fn}), we find that the single- and double-pole residues can be expressed as 
\begin{eqnarray*}
      f_1(e)  \!\!\!\!&=&\!\!\!\!
                          2 b_1 e^2 + b_2 e^4 + \fr{2}{3} b_3 e^6 + \cdots ,
                  \nonumber \\
      f_2(e)  \!\!\!\!&=&\!\!\!\!
                          - 2 b_1^2 e^4  -\fr{8}{3} b_2 b_1 e^6  + \cdots ,
                  \nonumber \\
      f_3(e)  \!\!\!\!&=&\!\!\!\!
                          \fr{8}{3} b_1^3 e^6  + \cdots .
\end{eqnarray*}

The metric field is set to $g_{\mu\nu}= e^{2\phi} \eta_{\mu\nu}$, as in the latter half of the previous section. Expanding the renormalization factor as $Z_3-1= \sum_{n=1}^\infty x_n(e)/(D-4)^n$ as usual, the QED bare action with field-strength $F_{0\mu\nu}=Z_3^{1/2} F_{\mu\nu}$ is expanded as follows: 
\begin{eqnarray}
   I_A  \!\!\!\!&=&\!\!\!\!
               - \fr{1}{4} \int  \! d^D x \sq{-g} \, F_{0\mu\nu} F_{0\lam\s} \, g^{\mu\lam} g^{\nu\s}
                   \nonumber \\
        \!\!\!\!&=&\!\!\!\!
               - \fr{1}{4} Z_3  \! \int  \! d^D x \, e^{(D-4)\phi} F_{\mu\nu} F_{\lam\s} 
                                 \eta^{\mu\lam} \eta^{\nu\s}
                   \nonumber \\
        \!\!\!\!&=&\!\!\!\!
               - \fr{1}{4} \int  \! d^D x  \, \biggl\{  \biggl( 1 + \fr{x_1(e)}{D-4} + \fr{x_2(e)}{(D-4)^2} 
                            + \cdots \biggr) F_{\mu\nu} F_{\lam\s} \eta^{\mu\lam} \eta^{\nu\s}
                   \nonumber \\
        \!\!\!\!&&\!\!\!\!   \quad
                 + \biggl( D-4 + x_1(e) + \fr{x_2(e)}{D-4} +  \cdots \biggr) 
                               \phi F_{\mu\nu} F_{\lam\s} \eta^{\mu\lam} \eta^{\nu\s}
                   \nonumber \\
        \!\!\!\!&&\!\!\!\!   \quad
                 + \half \bigl[ (D-4)^2 + (D-4) x_1(e) + x_2(e) +  \cdots \bigr] 
                               \phi^2 F_{\mu\nu} F_{\lam\s} \eta^{\mu\lam} \eta^{\nu\s}
                    \nonumber \\
       \!\!\!\!&&\!\!\!\!     \quad
                + \cdots \biggr\}
          \label{laurent expansion of qed action} .
\end{eqnarray}
The residue $x_n$ is given by $x_1=f_1$, $x_2=f_2+f_1^2/2$, $x_3 = f_3 + f_2 f_1 + f_1^3/6$ and so on using the previously defined $f_n$, thus
\begin{eqnarray}
   x_1(e)  \!\!\!\!&=&\!\!\!\! 
                          2 b_1e^2 + b_2 e^4 + \fr{2}{3} b_3 e^6 + \cdots,     
             \nonumber \\
   x_2(e)  \!\!\!\!&=&\!\!\!\!
                         - \fr{2}{3} b_2 b_1 e^6  + \cdots ,
        \label{residues of x_n}
\end{eqnarray}
and $x_3 = o(e^8)$. Note here that $o(e^4)$ of the residue $x_2$ and $o(e^6)$ of the residue $x_3$ disappear. This is a consequence of the gauge invariance, that is, $Z_e Z_3^{1/2}=1$, which represents that the 2- and 3-loop self-energy diagrams of the gauge field do not produce double and triple poles, respectively.

The first line on the right-hand side of (\ref{laurent expansion of qed action}) gives a normal kinetic term and counterterms of the gauge field. The second and following lines give new terms which do not appear in normal quantum field theory in the flat spacetime. Terms with negative power of $D-4$ are set as counterterms for eliminating UV divergences, and terms with zero or positive power are treated as new vertex functions.

A massless fermion field is conformally invariant in any dimension, and we can rescale the field appropriately to eliminate the $\phi$-dependence. Since the result is independent of how to choose the measure as mentioned before,  when calculating the effective action, use of the fermion action with the $\phi$-dependence removed simplifies the calculations.

Below, we calculate the effective action of QED in the case of $\phi \neq 0$ up to 3 loops, and confirm that momentum actually appears in the form of $Q$ (\ref{physical momentum}). First, we write down the normal effective action in Minkowski spacetime, which has been calculated as \cite{weinbergbook}
\begin{eqnarray*}
    \Gm_{\rm QED} \vert_{\phi=0}  \!\!\!\!&=&\!\!\!\!
             -\fr{1}{4} \biggl[ 1 -  \bigl( b_1 e^2 + b_2 e^4  +  b_3 e^6 \bigr) \log \fr{q^2}{\mu^2}
                        \nonumber \\
    \!\!\!\!&&\!\!\!\!     \qquad 
            -  \half b_2 b_1 e^6 \log^2 \fr{q^2}{\mu^2}      
                     \biggr] F_{\mu\nu} F_{\lam\s}(q) \eta^{\mu\lam} \eta^{\nu\s} .
\end{eqnarray*}
Here, only nonlocal terms are considered as quantum corrections.

There are various contributions to the part depending  on $\phi$, namely, the Wess-Zumino action. One is the finite terms derived from the Laurent expansion of the bare action (\ref{laurent expansion of qed action}), given by
\begin{eqnarray*}
     S(\phi)  \!\!\!\!&=&\!\!\!\! 
                   -\fr{1}{4} \biggl[  x_1(e) \phi  + \half x_2(e) \phi^2 \biggr] 
                   F_{\mu\nu} F_{\lam\s}(q)  \eta^{\mu\lam} \eta^{\nu\s}
               \nonumber \\
        \!\!\!\!&=&\!\!\!\!
                 -\fr{1}{4} \biggl[  \biggl( 2 b_1e^2 + b_2 e^4 + \fr{2}{3} b_3 e^6 \biggr) \phi  
                 - \fr{1}{3} b_2 b_1 e^6 \phi^2 \biggr] F_{\mu\nu} F^{\lam\s}(q)
                              \eta^{\mu\lam} \eta^{\nu\s}  .
\end{eqnarray*}
Others are contributions from finite loop corrections in Figs. 1 and 2. Fig. 1 has a single $\phi$ as an external field, and Fig. 2 has $\phi^2$, in which the solid, wavy, and dotted lines represent the $\phi$-field, gauge field, and fermion field, respectively, and diagrams that include counterterms as subdiagrams are not depicted here. These figures are obtained by inserting the vertex functions containing $\phi$ into the 2- and 3-loop self-energy diagrams of the gauge field. Calculations are performed with momentum of $\phi$ as zero. They are easily achieved, and adding all these contributions results in 
\begin{eqnarray}
    V(\phi)  \!\!\!\!&=&\!\!\!\!
                     -\fr{1}{4} \biggl[  \biggl( b_2 e^4  +  \fr{4}{3} b_3 e^6  
                     + 2b_2 b_1 e^6 \log \fr{q^2}{\mu^2} \biggr)\phi  
                \nonumber \\
      \!\!\!\!&&\!\!\!\!     \qquad
               - \fr{5}{3} b_2 b_1 e^6 \phi^2  \biggr] F_{\mu\nu} F_{\lam\s}(q)
                                  \eta^{\mu\lam} \eta^{\nu\s} .
          \label{loop contribution v}
\end{eqnarray}
Details of each term will be described below.

First, we write out the results from the 2-loop self-energy (2LSE) and 3-loop self-energy (3LSE) diagrams of the gauge field required in the following calculations of $V$ (\ref{loop contribution v}), which are given by 
\begin{eqnarray}
     \hbox{2LSE} = \fr{x_1^{(4)}}{2} \biggl( \fr{1}{\eps} -2 \log \fr{q^2}{\mu^2} \biggr)
                             \biggl( -\fr{1}{4} F^2\biggr) 
            \label{2LSE}
\end{eqnarray}
and
\begin{eqnarray}
    \hbox{3LSE} = \biggl( - \fr{x_2^{(6)}}{4}  \fr{1}{\eps^2} 
                                     + \fr{x_1^{(6)}}{2} \fr{1}{\eps}  \biggr)
                             \biggl( -\fr{1}{4} F^2 \biggr) ,
          \label{3LSE}
\end{eqnarray}
where $F^2$ is an abbreviation for $F_{\mu\nu} F_{\lam\s}(q) \eta^{\mu\lam} \eta^{\nu\s}$ and the calculation is performed by setting the dimension as $D=4-2\eps$. The $o(e^m)$ component of the $n$th-pole residue $x_n$ is denoted as $x_n^{(m)}$, and each coefficient is read from (\ref{residues of x_n}) as $x_1^{(2)}=2b_1 e^2$, $x_1^{(4)}=b_2 e^4$, $x_1^{(6)}=2b_3 e^6/3$, $x_2^{(6)}=-2b_2 b_1 e^6/3$. The counterterms in the first line of (\ref{laurent expansion of qed action}) are designed to eliminate these UV divergences. As mentioned before, 2LSE has no double poles and also 3LSE has no triple poles. For 3LSE, there are diagrams containing one fermion loop and two fermion loops, and the double pole arises only from diagrams with two fermion loops. Moreover, all divergences are local due to renormalizablility, that is, nonlocal ones such as $(1/\eps) \times \log (q^2/\mu^2)$ do not appear.

\begin{figure}[h]
\begin{center}
\includegraphics[width=16cm]{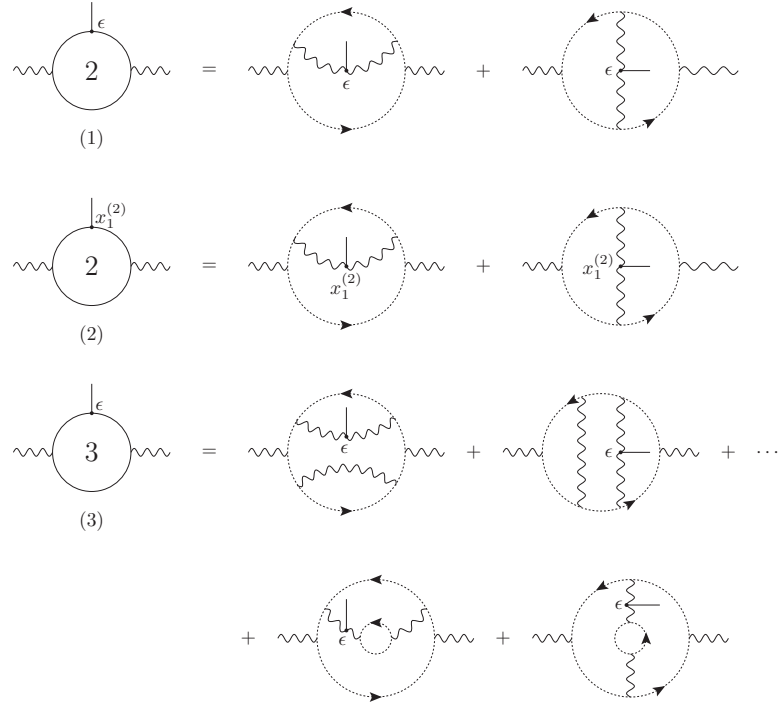}
\end{center}
\vspace{-10.5cm}
\caption{Finite vertex contributions with a single $\phi$ and two external background gauge fields.}
\end{figure}

Fig. 1-(1) shows the insertion of the vertex function $-2\eps \phi \, (-F^2/4)$ into an internal line of the gauge field in the 2LSE diagram, so that the simple pole of (\ref{2LSE}) and $\eps$ in the vertex function cancel out to be finite. It gives the first $o(e^4)$ term of (\ref{loop contribution v}). Fig. 1-(2) is a 2LSE diagram with the vertex function $x_1^{(2)} \phi \, (-F^2/4)$ inserted. From this diagram, the third nonlocal term with a single $\phi$ is obtained as a finite term. At the same time, a simple-pole divergence $(- x_1^{(2)} x_1^{(4)}/2)(1/\eps) \, \phi \, (-F^2/4)$ occurs. Fig. 1-(3) shows a 3LSE diagram with the vertex function $-2\eps \phi \, (-F^2/4)$ inserted. The $\eps$ at the vertex acts to the double pole of 3LSE, generating a simple-pole divergence $-x_2^{(6)} (1/\eps) \, \phi \, (-F^2/4)$, and also cancels out the simple pole, resulting in a finite term $2x_1^{(6)} \phi \, (-F^2/4)$, which is the local $o(e^6)$ term with a single $\phi$ in  (\ref{loop contribution v}). Here, the sum of the divergences from Fig. 1-(2) and Fig. 1-(3) cancels out the counterterm $(-x_2/2)(1/\eps) \, \phi \, ( -F^2/4)$ in the bare action (\ref{laurent expansion of qed action}), which can be seen from the result that the sum of the coefficients, $-x_1^{(2)} x_1^{(4)}/2-x_2^{(6)} = -b_2 b_1 e^6/3 = x_2^{(6)}/2$, has the opposite sign of it.

\begin{figure}[h]
\begin{center}
\includegraphics[width=16cm]{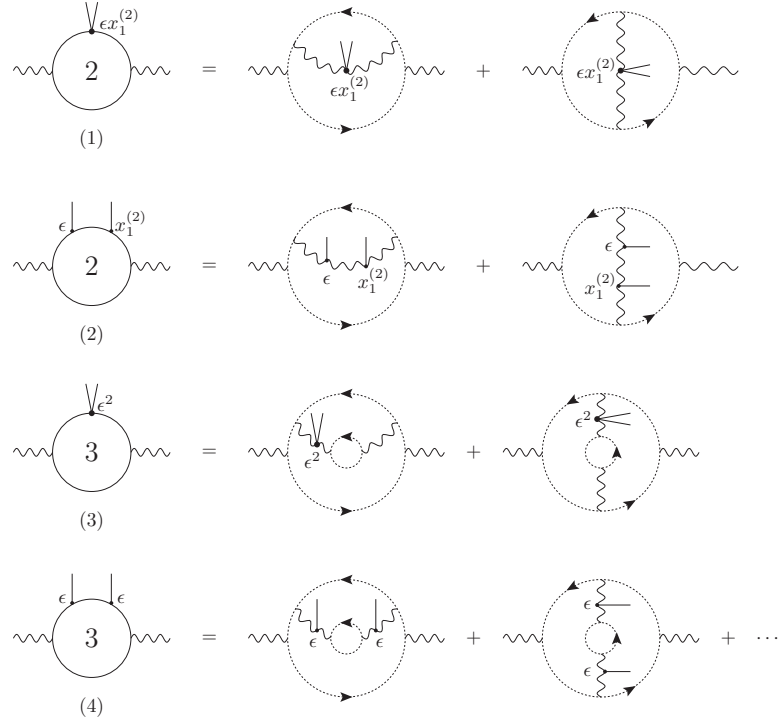}
\end{center}
\vspace{-10.5cm}
\caption{Finite vertex contributions with two $\phi$ and two external background gauge fields.}
\end{figure}

The last term in (\ref{loop contribution v}) with $\phi^2$ is derived by summing all finite contributions in Fig. 2. Fig. 2-(1) represents a 2LSE diagram with the vertex function $-\eps x_1^{(2)} \phi^2 (-F^2/4)$ inserted, resulting in a finite contribution $(x_1^{(2 )} x_1^{(4)}/2) \, \phi^2 (-F^2/4)$. Fig. 2-(2) is a 2LSE diagram with one $-2\eps \phi \, (-F^2/4)$ and one $x_1^{(2)} \phi \, (-F^2/4)$ inserted, which gives a finite $-2 x_1^{(2)} x_1^{(4)} \phi^2 \, (-F^2/4)$. Fig. 2-(3) is obtained by inserting $2\eps^2 \phi^2 (-F^2/4)$ into one of the two internal gauge field lines in the 3LSE diagram with double poles, which generates a finite $x_2^{(6)} \phi^2 \, (-F^2/4)$. Fig. 2-(4) is obtained by inserting two $-2\eps \phi (-F^2/4)$ into two internal gauge field lines in 3LSE with double poles, where the dots denote other variations of how to insert the vertex. This gives a finite contribution $-3x_2^{(6)} \phi^2 (-F^2/4)$. Adding these yields the last term of (\ref{loop contribution v}) with coefficient $x_1^{(2)} x_1^{(4)}/2 -2 x_1^{(2)} x_1^{(4)} + x_2^{(6)}- 3x_2^{(6)} = (- 5/3) b_2 b_1 e^6$.

Adding all the contributions, we find that the effective action can be written in terms of the physical momentum (\ref{physical momentum}) as follows:
\begin{eqnarray*}
   \Gm_{\rm QED}(g)   \!\!\!\!&=&\!\!\!\!
             \Gm_{\rm QED} \vert_{\phi=0} + S(\phi) + V(\phi)
                       \nonumber \\ 
     \!\!\!\!&=&\!\!\!\!
           -\fr{1}{4} \biggl[ 1 -  \bigl( b_1 e^2  + b_2 e^4  + b_3 e^6 \bigr) \log \fr{Q^2}{\mu^2}
                        \nonumber \\
     \!\!\!\!&&\!\!\!\!   \qquad 
           -  \half b_2 b_1 e^6 \log^2 \fr{Q^2}{\mu^2}  
                   \biggr] F_{\mu\nu} F_{\lam\s}(q) \eta^{\mu\lam} \eta^{\nu\s} .
\end{eqnarray*}
In this way, it is confirmed by the direct calculations that the effective action in a conformally flat spacetime is indeed constructed in terms of (\ref{final diffeomorphism invariant form}). The effective action in any curved spacetime can be inferred from diffeomorphism invariance and gauge invariance.

\section{Effective Actions of QCD and Quantum Gravity}
\setcounter{equation}{0}
\noindent

Since the basic structure of the Wess-Zumino action for conformal anomaly does not depend on the gauge group, the same holds for QCD, or Yang-Mills theory, as for QED. Using dimensional regularization, the bare action of the Yang-Mills gauge field is also expanded like (\ref{laurent expansion of qed action}), thus the effective action can be calculated in the same way. Since diffeomorphism invariance is guaranteed, it will be expressed in terms of the physical momentum $Q$ (\ref{physical momentum}).

Therefore, let $\geh$ be a coupling constant of QCD and its beta function be 
\begin{eqnarray*}
   \b_\geh = \mu \fr{d \geh}{d \mu} = - \b_0 \geh^3 - \b_1 \geh^5 - \b_2 \geh^7 - \cdots ,
\end{eqnarray*}
then the effective action will be given by\footnote{
\label{footnote on background field method}The background field method \cite{abbott} is convenient for calculating a gauge-invariant effective action. When normalizing the gauge field as usual so that the kinetic term does not depend on the coupling constant, a renormalization factor of the coupling constant, $Z_\geh$, and that of the background gauge field, $Z_A^{1/2}$, satisfy the same identity $Z_\geh Z_A^{1/2}=1$ as in QED. This also represents that the background gauge field normalized as in (\ref{qcd effective action}) does not receive renormalization.} 
\begin{eqnarray}
  \Gm_{\rm QCD}   \!\!\!\!&=&\!\!\!\!
           - \fr{1}{4} \int \! \fr{d^4 q}{(2\pi)^4} \biggl[   \fr{1}{\geh^2} 
           + \bigl( \b_0 + \b_1 \geh^2 + \b_2 \geh^4 \bigr) \log \biggl( \fr{Q^2}{\mu^2} \biggr)
                   \nonumber \\
      \!\!\!\!&&\!\!\!\!     \qquad\qquad
             - \half \b_1 \b_0 \geh^4 \log^2 \biggl( \fr{Q^2}{\mu^2} \biggr)  
            +  \cdots   \biggr]   
             F^a_{\mu\nu}F^a_{\lam\s} (q) \, \eta^{\mu\lam} \eta^{\nu\s} .
             \label{qcd effective action}
\end{eqnarray}
The difference from the previous section is that the beta function is negative and that the gauge field is normalized to factor out $1/\geh^2$ for the following discussion.

In the case of QCD, we can introduce a running coupling constant that becomes small at high energy. It can be expressed as
\begin{eqnarray*}
   \bar{\geh}^2 (Q)  \!\!\!\!&=&\!\!\!\! 
          \fr{1}{\b_0 \log \fr{Q^2}{\Lam_\QCD^2}} \, \Biggl\{  1 - 
    \fr{\b_1 \log \Bigl( \log \fr{Q^2}{\Lam_\QCD^2} \Bigr)}{\b_0^2 \log \fr{Q^2}{\Lam_\QCD^2}}  
          + \fr{\b_1^2}{\b_0^4 \log^2 \fr{Q^2}{\Lam_\QCD^2}} 
           \biggl[  \log^2 \biggl( \log \fr{Q^2}{\Lam_\QCD^2} \biggr) 
          \nonumber \\
     && \qquad\qquad\qquad
              - \log \biggl( \log \fr{Q^2}{\Lam_\QCD^2} \biggr)    
              + \fr{\b_2 \b_0}{\b_1^2} - 1 \biggr]   + \cdots    \Biggr\} 
\end{eqnarray*}
for $Q \gg \Lam_{\rm QCD}$, where $\Lam_{\rm QCD}$ is a physical energy scale of QCD. Rewriting the effective action (\ref{qcd effective action}) using the running coupling constant results in the simple form
\begin{eqnarray}
  \Gm_{\rm QCD}  = -\fr{1}{4} \int \! \fr{d^4 q}{(2\pi)^4} \, \fr{1}{\bar{\geh}^2 (Q)} \,
               F^a_{\mu\nu}F^a_{\lam\s} (q) \, \eta^{\mu\lam} \eta^{\nu\s} .
       \label{final form of qcd effective action}
\end{eqnarray}
This reflects the fact that the effective action is a RG invariant. For details, see Appendix. This expression infers that when the physical momentum $Q$ becomes less than $\Lam_{\rm QCD}$ and $\bar{\geh}(Q)$ diverges, the kinetic term of the gauge field vanishes, that is, the dynamics disappear and confinement will occur.

The conformal anomaly composed of the gravitational field is also determined from diffeomorphism invariance, thus the form of relevant quantities is basically unchanged apart from the coefficient whether the gravitational field is quantized or not. The essential difference between them will be described last from a viewpoint of symmetry based on recent research.

From studies of the conformal anomalies applying Hathrell's RG method to QED \cite{hathrellQED} and QCD \cite{freeman} in curved spacetime, it has been shown that at least in these theories, gravitational part of the conformal anomalies are classified into only two \cite{hamada14CS, hm16}: the Weyl tensor squared,
\begin{eqnarray}
    C_{\mu\nu\lam\s}^2 = R_{\mu\nu\lam\s}^2 - 2 R_{\mu\nu}^2 + \fr{1}{3} R^2 ,
       \label{weyl tensor squared}
\end{eqnarray}
and the Euler density predicted by Riegert \cite{riegert},
\begin{eqnarray}
     E_4 = G_4 - \fr{2}{3} \nb^2 R ,
        \label{modified Euler density}
\end{eqnarray}
where $G_4=R_{\mu\nu\lam\s}^2 -4R_{\mu\nu}^2 +R^2$ is the normal Euler density. Since the modification is total-divergence, the volume integral of $E_4$ is the same as that of $G_4$. The corresponding gravitational action is given by 
\begin{eqnarray}
    I_G = \int d^4 x \sq{-g} \biggl[  - \fr{1}{t^2} C^2_{\mu\nu\lam\s}  - b G_4 \biggr] .
         \label{conformal gravity action}
\end{eqnarray}
Here, $t$ is a coupling constant that controls dynamics of traceless tensor fields in  renormalizable quantum conformal gravity \cite{hs, hamada02, hamada14, hm16, hm17, book}. On the other hand, since the Euler term does not have a kinetic term of the gravitational field, $b$ is not an independent coupling constant, but it gives a pure counterterm for eliminating divergences proportional to $G_4$.\footnote{
In dimensional regularization, the bare $b$ (omitting the subscript $0$) is expanded by a pure-pole factor like $\sum_{n=1}^\infty b_n/(D-4)^n$, and $G_4$ is extended to a $D$-dimensional quantity $G_D=G_4 +(D-4)\chi(D)H^2$, where $H=R/(D-1)$ and $\chi(D)=\sum_{n=1}^\infty \chi_n (D-4)^{n-1}$ \cite{hamada14CS, hamada14, hm16}. The coefficient $\chi_n$ can be determined in order by solving the Hathrell's RG equations, and first three have been calculated as $\chi_1 =1/2$, $\chi_2=3/4$, and $\chi_3=1/3$. The Riegert action and its series, as discussed later, are generated by Laurent-expanding a bare action $b G_D$.} 

The Weyl tensor squared (\ref{weyl tensor squared}) satisfies $\sq{-g} \, C^2_{\mu\nu\lam\s}=\sq{-\bg} \, {\bar C}_{\mu\nu\lam\s}^2$ in 4 dimensions when the metric field is decomposed as in (\ref{conformal metric decomposition}), where the curvature with the bar is that composed of the metric $\bg_{\mu\nu}$. Furthermore, $\bg_{\mu\nu}$ is expanded by introducing the traceless tensor field $h^\mu_{~\nu}$ as $\bg_{\mu\nu}= (\hg e^{h} )_{\mu\nu} = \hg_{\mu\lam} ( \dl^\lam_{~\nu} + h^\lam_{~\nu} + \cdots)$, where $\hg_{\mu\nu}$ is a background metric. If the background spacetime is taken to be the flat and the field is normalized as $h^\mu_{~\nu} \to th^\mu_{~\nu}$, then the Weyl action in $I_G$ (\ref{conformal gravity action}) that gives the kinetic term of the traceless tensor field is expanded as $\int d^4x [ - \pd^2 h^\mu_{~\nu} \pd^2 h^\nu_{~\mu} /2 + \pd^\mu \chi^\nu \pd_\mu \chi_\nu  - \pd_\mu \chi^\mu \pd_\nu \chi^\nu/3] + o(th^3)$, where $\chi_\mu = \pd_\nu h^\nu_{~\mu}$ and the indices are contracted by $\eta_{\mu\nu}$. Then, the effective action of the quantum gravity can be calculated as an ordinary quantum field theory defined in the flat spacetime.\footnote{
As in footnote \ref{footnote on background field method}, if the effective action is calculated by introducing the background traceless tensor field $\hat{h}^\mu_{~\nu}$ as $\hg_{\mu\nu} =(e^{t{\hat h}})^\mu_{~\nu}$, then diffeomorphism invariance demands that a renormalization factor of the field, $Z_{\hat h}^{1/2}$, and that of the coupling constant, $Z_t$, satisfy the identity $Z_t Z_{\hat h}^{1/2}=1$. Diffeomorphism invariance also demands that a nonrenomalization theorem holds for $\phi$ because this field is treated exactly without introducing a coupling constant for it \cite{hamada02, hamada14, hm16, hm17, book}.} 

The Weyl part has a structure similar to the gauge field. Now, ${\cal L}^{(0)}(g)$ is $\sq{-g} \, C^2_{\mu\nu\lam\s}$, thus the first Wess-Zumino action is given by
\begin{eqnarray*}
     S_W^{(1)}(\phi, \bg)  \!\!\!\!&=&\!\!\!\!
               \int d^4 x \int^\phi_0 d\phi \sq{-g} \, C^2_{\mu\nu\lam\s}
                   \nonumber \\
       \!\!\!\!&=&\!\!\!\!
               \int d^4 x \sq{-g} \, \phi \, C^2_{\mu\nu\lam\s} .
\end{eqnarray*}
Further integrating it, we obtain $S^{(n)}_{W \, {\rm top}}= (1/n!) \times \int d^4 x \sq{-g} \, \phi^n C^2_{\mu\nu\lam\s}$ that satisfies the recursion relation (\ref{reduction formula for top term}), and then ${\cal L}^{(n)}(g)$, which is a covariant form containing it, will be given by $\log^n (Q^2 /\mu^2) \sq{-g} \, C^2_{\mu\nu\lam\s}$ in momentum space, as in (\ref{final diffeomorphism invariant form}). Since the beta function of the coupling constant $t$ is known to be negative, we can express the logarithmic part using a running coupling constant $\bar{t}^2(Q)$ as in QCD, which is obtained by replacing $\geh$ with $t$ and, accordingly, renaming $\Lam_{\rm QCD}$ to $\Lam_{\rm QG}$, then the effective action will be given by replacing $t^2$ in (\ref{conformal gravity action}) with $\bar{t}^2(Q)$ \cite{hamada02, hamada14, book}.\footnote{
Note that in dimensional regularization, a $D$-dimensional Weyl tensor squared has the structure $\sq{-g} \,C_{\mu\nu\lam\s}^2 = \sq{-\bg} \, e^{(D-4)\phi} \, {\bar C}_{\mu\nu\lam\s}^2$ similar to the gauge field squared.} 
This suggests that when $\bar{t}(Q)$ becomes larger at the new physical scale $\Lam_{\rm QG}$, the 4th-derivative conformal gravity dynamics disappears.

Next, we consider the Wess-Zumino action for the modified Euler density (\ref{modified Euler density}). In this case, if the argument is made with $\phi$ as a constant as before, the essence will be lacking, so it is treated as a function in coordinate space here. Its first Wess-Zumino action is then given by \cite{riegert}
\begin{eqnarray}
      S_R^{(1)} (\phi,\bg)  \!\!\!\!&=&\!\!\!\!
                  \int d^4 x \int^\phi_0 d\phi \sq{-g} \, E_4
                      \nonumber \\
         \!\!\!\!&=&\!\!\!\!
                  \int d^4 x \sq{-\bg} \, \bigl( 2 \phi {\bar \Delta}_4 \, \phi 
                                               + {\bar E}_4 \, \phi \bigr) .
           \label{riegert action}
\end{eqnarray}
The integration of $\phi$ is performed using the relation $\sq{-g} \, E_4 = \sq{-\bg} \, (4 {\bar \Delta}_4 \phi + { \bar E}_4) $, where $\sq{-g}\Delta_4$ is a conformally invariant differential operator defined by
\begin{eqnarray*}
     \Delta_4 = \nb^4  + 2 R^{\mu\nu} \nb_\mu \nb_\nu - \fr{2}{3} R \nb^2 
                    + \fr{1}{3} \nb^\mu R \nb_\mu ,
\end{eqnarray*}
which satisfies $\sq{-g} \, \Delta_4 A = \sq{-\bg} \, {\bar \Delta}_4 A$ and a self-adjointness $\int d^4x \sq{-g} A \Delta_4 B = \int d^4 x \sq{-g} (\Delta_4 A) B$ for for any scalar $A$ and $B$. The action (\ref{riegert action}), called the Riegert action, works as a kinetic term of the conformal-factor field \cite{am, amm92, amm97, hs}.

The first Wess-Zumino action can be written in a diffeomorphism invariant form as
\begin{eqnarray*}
      \Gm_R^{(1)}(g)  \!\!\!\!&=&\!\!\!\!   
                S^{(1)}_R (\phi, \bg) + \Gm^{(1)}_R (\bg) 
                    \nonumber \\
         \!\!\!\!&=&\!\!\!\!
                \fr{1}{8} \int d^4 x \sq{-g} E_4 \fr{1}{\Delta_4} E_4 ,
\end{eqnarray*}
where $\Delta_4^{-1} E_4 (x) = \int d^4 y  \, G(x,y) \sq{-g} \, E_4(y)$ and $\Delta_4 G(x,y)=\dl^4 (x-y)/\sq{-g}$. This action has a nonlocal structure, but does not contain the scale $\mu$. Also, unlike other Wess-Zumino actions, it appears even in the zeroth order of the coupling constant.

Further integrating the first Wess-Zumino action yields 
\begin{eqnarray*}
   S^{(n)}_{R \, {\rm top}} (\phi,\bg) = \fr{1}{n!} \int d^4 x \sq{-\bg} \bigl( 2 \phi^n {\bar \Delta}_4 \phi + {\bar E}_4 \phi^n \bigr) .
\end{eqnarray*}
Using the self-adjointness of ${\bar \Delta}_4$ and $\int d^4 x \sq{-\bg} \, {\bar \Delta}_4 A =0$, we can show that this action satisfies the recursion relation (\ref{reduction formula for top term}).

Now, we consider a diffeomorphism invariant effective action containing $S^{(n)}_{R \, {\rm top}}$, which will appear in higher-loop corrections. To begin with, $\Gm^{(1)}_R$ is a quantity that appears even in the zeroth order of the coupling constant, thus it does not involved the scale $\mu$, as shown above. However, if corrections by the coupling constant $t$ are added, the logarithmic correction will be accompanied. From these considerations and paying attention to $\log \Delta_4= - 4\phi  + \log {\bar \Delta}_4$, we deduce the following diffeomorphism invariant action that contains $S^{(n)}_{R \, {\rm top}}$ as a local part:
\begin{eqnarray}
     \Gm_R^{(n)} (g) =  \fr{1}{8} \fr{(-1)^{n-1}}{4^{n-1}n!} \int d^4 x \sq{-g} E_4 \log^{n-1} \biggl( \fr{\Delta_4}{\mu^4} \biggr)
                   \fr{1}{\Delta_4} E_4 .
       \label{logarithmic riegert action}
\end{eqnarray}
Here the logarithmic part suggests that the effective action will be written in terms of the running coupling constant $\bar{t}(Q)$.

In quantum theory of gravity, the total energy-momentum tensor must vanish as an equation of motion of quantum gravitational fields.  It implies that conformal invariance is realized as diffeomorphism invariance at the quantum level. The Riegert action (\ref{riegert action}) then arises rather as necessary to restore the conformal invariance. This conformal symmetry is a gauge symmetry that appears only when gravity is quantized.\footnote{
Here note that the traceless tensor field is controlled by perturbation, whereas the conformal-factor field is handled exactly as in (\ref{conformal metric decomposition}). This treatment is significant when constructing renormalizable quantum gravity with this conformal invariance asymptotically \cite{hs, hamada02, hamada14, hm16, hm17, book}.} 
It means that all theories with different backgrounds connected to each other by conformal transformations are gauge equivalent, that is, background-metric independent. Whether it exists or not is the difference between quantum gravity and quantum field theory in curved spacetime. It has been shown that under this symmetry, all of ghost modes become gauge variant, namely, unphysical, even at the UV limit, thus they are confined in quantum spacetime \cite{hh, hamada12M4, hamada12RxS3, book}.

\section{Conclusion}
\setcounter{equation}{0}
\noindent

In this paper, we examined the Wess-Zumino actions at higher loops that are obtained by integrating the conformal anomalies with respect to the conformal-factor field. The consistency condition (\ref{wess-zumino condition}) that the Wess-Zumino actions should satisfy was derived from the fact that the effective action is diffeomorphism invariant, and a series of the Wess-Zumino actions satisfying the condition was constructed by repeating the integration. We have seen that they arise to make the nonlocal loop correction terms diffeomorphism invariant and that the effective action is described in terms of the physical momentum (\ref{physical momentum}). Thus, conformal anomalies are indispensable quantities to preserve diffeomorphism invariance, so that we must always incorporate them when considering quantum field theory in curved spacetime and quantum gravity.

As a specific example, the QED effective action in a conformally flat spacetime was calculated at the 3-loop level. It was done by employing dimensional regularization that obviously preserves both gauge and diffeomorphism invariances. In this method, the result does not depend on the choice of the path integral measure. Instead, the information of conformal anomalies is included between $D$ and 4 dimensions, thus it requires careful handling of finite quantities that appear when the poles and zeros of $D-4$ cancel out.  In this way, we confirmed that a series of the Wess-Zumino actions is actually realized.

The same will hold true for the effective action of QCD from gauge symmetry and diffeomorphism invariance. Here we have shown that it can be summarized in the form of the reciprocal of the running coupling constant squared described in the physical momentum. We also examined the effective action of renormalizable quantum conformal gravity. The Wess-Zumino action obtained by integrating the Weyl tensor squared has the same structure as that of the gauge field, thus the Weyl part of the effective action will be written in terms of the running coupling constant as well. Furthermore, we considered a series of the Wess-Zumino actions concerning the Euler density, and deduced that the corresponding effective action will be given by the Riegert action with logarithmic nonlocality (\ref{logarithmic riegert action}) that can be regarded as a realization of the running coupling constant.

Finally, in QCD and the quantum gravity, we have seen that the effective action will be summarized in a form in which the kinetic term disappears when the running coupling constant diverges, as in (\ref{final form of qcd effective action}). Under this consideration, we can construct a simplified model of strong coupling dynamics occurring at low energy by approximating the running coupling constant as a coordinate-dependent average. Applying such a mean field approximation to the quantum gravity, we can describe a spacetime transition that the conformal gravity dynamics disappear at the energy scale $\Lam_{\rm QG}$ predicted to be about $10^{17}$ GeV, below the Planck scale, and shift to Einstein gravity \cite{hhy, hamada20, book}. One of the aim of this paper is to theoretically reinforce those achievements.

\appendix

\section{Running Coupling Constant and Effective Action}
\setcounter{equation}{0}
\noindent

In this appendix, we show the relationship between  the effective action and the running coupling constant. Here, more generally, the beta function is expanded up to
\begin{eqnarray}
   \b_\geh = \mu \fr{d \geh}{d \mu} 
       = - \b_0 \geh^3 - \b_1 \geh^5 - \b_2 \geh^7 -\b_3 \geh^9 
          \label{qcd beta function}
\end{eqnarray}
and higher-order coefficients are taken to be zero.

Integrating (\ref{qcd beta function}) gives
\begin{eqnarray*}
    \log \mu + C = \int^\geh_{\bar{\geh}(C)} \fr{d \geh^\pp}{\b_\geh(\geh^\pp)} .
\end{eqnarray*}
The lower limit of integration, $\bar{\geh}(C)$, is a constant, that is, a RG invariant satisfying $\mu d \bar{\geh}(C)/d\mu=0$. Here, it is decided so that the integration constant $C$ becomes $-\log Q$. If we write it as $\bar{\geh}(Q)$ and let it be a function that satisfies $\bar{\geh}(\mu)=\geh$, then 
\begin{eqnarray*}
     - \log \fr{Q}{\mu}  =  \int^\geh_{\bar{\geh}(Q)} \fr{d\geh^\pp}{\b_\geh(\geh^\pp)}
                       = \half \int^{\bar{\geh}^2(Q)}_{\geh^2} \fr{d\lam}{\lam^2} 
                            \fr{1}{\b_0 + \b_1 \lam + \b_2 \lam^2 + \b_3 \lam^3}
\end{eqnarray*}
is obtained. The function $\bar{\geh}(Q)$ is called the running coupling constant.

Performing the integration up to the 6th order yields
\begin{eqnarray}
      && \fr{1}{\bar{\geh}^2(Q)} + \fr{\b_1}{\b_0} \log \bigl[ \b_0 \bar{\geh}^2(Q) \bigr]
          + \biggl( \fr{\b_2}{\b_0} - \fr{\b_1^2}{\b_0^2} \biggr) \bar{\geh}^2(Q) 
          + \half \biggl( \fr{\b_3}{\b_0} - \fr{2 \b_2 \b_1}{\b_0^2} 
                \nonumber \\
      && \qquad 
            + \fr{\b_1^3}{\b_0^3} \biggr) \bar{\geh}^4(Q)
            - \fr{1}{3} \biggl( \fr{2 \b_3 \b_1}{\b_0^2} + \fr{\b_2^2}{\b_0^2} 
                    - \fr{3 \b_2 \b_1^2}{\b_0^3} + \fr{\b_1^4}{\b_0^4} \biggr) \bar{\geh}^6(Q)
               \nonumber \\
     && = \b_0 \log \fr{Q^2}{\Lam_\QCD^2} ,
            \label{defining equation of running coupling}
\end{eqnarray}
where the right-hand side is defined by
\begin{eqnarray*}
   &&  \b_0 \log \fr{Q^2}{\Lam_{\rm QCD}^2}  \equiv \b_0 \log \fr{Q^2}{\mu^2 } 
          + \fr{1}{\geh^2} + \fr{\b_1}{\b_0} \log \bigl( \b_0 \geh^2 \bigr)
          + \biggl( \fr{\b_2}{\b_0} - \fr{\b_1^2}{\b_0^2} \biggr) \geh^2 
                \nonumber \\
   &&  \qquad
             + \half \biggl( \fr{\b_3}{\b_0} - \fr{2 \b_2 \b_1}{\b_0^2} 
                             + \fr{\b_1^3}{\b_0^3} \biggr) \geh^4
             - \fr{1}{3} \biggl( \fr{2 \b_3 \b_1}{\b_0^2} + \fr{\b_2^2}{\b_0^2} 
                    - \fr{3 \b_2 \b_1^2}{\b_0^3} + \fr{\b_1^4}{\b_0^4} \biggr) \geh^6     
\end{eqnarray*}
and the constant $(\b_1/\b_0) \log \b_0$ has been added to both sides for convention. Since the left-hand side of (\ref{defining equation of running coupling}) is a RG invariant, the energy scale $\Lam_{\rm QCD}$ defined through this equation is also so, namely, it is a physical scale.

Let (\ref{defining equation of running coupling}) solve for $\bar{\geh}^2(Q)$ iteratively as $L=\log (Q^2 /\Lam^2_\QCD) \gg 1$, then the running coupling constant can be expressed as
\begin{eqnarray*}
   \bar{\geh}^2 (Q)  \!\!\!\!&=&\!\!\!\! 
         \fr{1}{\b_0 L} \Biggl\{ 1 - \fr{\b_1}{\b_0^2 L} \log L 
         + \fr{\b_1^2}{\b_0^4 L^2} \biggl( \log^2 L - \log L + \fr{\b_2 \b_0}{\b_1^2} -1 \biggr)
              \nonumber \\
      \!\!\!\!&&\!\!\!\!   \qquad
             - \fr{\b_1^3}{\b_0^6 L^3} \biggl[  \log^3 L - \fr{5}{2} \log^2 L 
             + \biggl( \fr{3\b_2 \b_0}{\b_1^2} -2 \biggr) \log L  - \fr{\b_3 \b_0^2}{2 \b_1^3}
             +\half \biggr] 
              \nonumber \\
     \!\!\!\!&&\!\!\!\!    \qquad 
             + \fr{\b_1^4}{\b_0^8 L^4} \biggl[ \log^4 L  - \fr{13}{3} \log^3 L
             + \biggl( \fr{6 \b_2 \b_0}{\b_1^2} - \fr{3}{2} \biggr) \log^2 L 
              - \biggl( \fr{2 \b_3 \b_0^2}{\b_1^3}
             \nonumber \\
      \!\!\!\!&&\!\!\!\!    \qquad\qquad\qquad
            + \fr{3 \b_2 \b_0}{\b_1^2} - 4 \biggr) \log L 
            - \fr{\b_3 \b_0^2}{6 \b_1^3} + \fr{5 \b_2^2 \b_0^2}{3 \b_1^4} 
            - \fr{3 \b_2 \b_0}{\b_1^2} + \fr{7}{6} \biggr]
           \Biggr\} .
\end{eqnarray*}
From this expression, we find that the effective action (\ref{final form of qcd effective action}) can be written as follows:
\begin{eqnarray*}
  \Gm_{\rm QCD}  \!\!\!\!&=&\!\!\!\!
                -\fr{1}{4} \int \! \fr{d^4 q}{(2\pi)^4} \, \fr{1}{\bar{\geh}^2 (Q)} \,
                F^a_{\mu\nu}F^a_{\lam\s} (q) \, \eta^{\mu\lam} \eta^{\nu\s} 
                   \nonumber \\
       \!\!\!\!&=&\!\!\!\!
                -\fr{1}{4} \int \! \fr{d^4 q}{(2\pi)^4} \,  \Biggl\{ \b_0 L + \fr{\b_1}{\b_0} \log L 
                + \fr{\b_1^2}{\b_0^3 L} \biggl( \log L - \fr{\b_2 \b_0}{\b_1^2} + 1 \biggr)
                    \nonumber \\
        \!\!\!\!&&\!\!\!\!    \qquad
                + \fr{\b_1^3}{\b_0^5 L^2} \biggl( - \half \log^2 L
                + \fr{\b_2 \b_0}{\b_1^2} \log L 
                - \fr{\b_3 \b_0^2}{2 \b_1^3}  + \half  \biggr)  
                    \nonumber \\  
        \!\!\!\!&&\!\!\!\!    \qquad
                + \fr{\b_1^4}{\b_0^7 L^3} \biggl[ \fr{1}{3} \log^3 L 
                - \biggl( \fr{\b_2 \b_0}{\b_1^2} + \half \biggr) \log^2 L 
                + \biggl( \fr{\b_3 \b_0^2}{\b_1^3}   + \fr{\b_2 \b_0}{\b_1^2} 
                   \nonumber \\
        \!\!\!\!&&\!\!\!\!    \qquad
                - 1 \biggr) \log L
                + \fr{\b_3 \b_0^2}{6 \b_1^3}
                - \fr{2 \b_2^2 \b_0^2}{3 \b_1^4} 
                + \fr{\b_2 \b_0}{\b_1^2} - \fr{1}{6} \biggr]  
                \Biggl\}  F^a_{\mu\nu}F^a_{\lam\s} (q) \, \eta^{\mu\lam} \eta^{\nu\s} 
                  \nonumber \\
        \!\!\!\!&=&\!\!\!\!
               -\fr{1}{4} \int \! \fr{d^4 q}{(2\pi)^4} \, \Biggl\{ 
               \fr{1}{\geh^2} + \bigl( \b_0 + \b_1 \geh^2 + \b_2 \geh^4 + \b_3 \geh^6 \bigr) 
               \log \biggl( \fr{Q^2}{\mu^2} \biggr)
                     \nonumber \\
        \!\!\!\!&&\!\!\!\!    \qquad
               - \half \Bigl[ \b_1 \b_0 \geh^4 + \bigl( 2 \b_2 \b_0 + \b_1^2 \bigr) \geh^6  \Bigr]
                    \log^2 \biggl( \fr{Q^2}{\mu^2} \biggr) 
                      \nonumber \\
        \!\!\!\!&&\!\!\!\!     \qquad  
               + \fr{1}{3} \b_1 \b_0^2 \geh^6 \log^3 \biggl( \fr{Q^2}{\mu^2} \biggr)  \Biggl\}  
                 F^a_{\mu\nu}F^a_{\lam\s} (q) \, \eta^{\mu\lam} \eta^{\nu\s} .
\end{eqnarray*}
Here, we consider the effective action up to $o(\geh^6)$. Seeing up to $o(\geh^4)$, it agrees with (\ref{qcd effective action}).

\end{document}